# Catalogue of known Galactic SNRs uncovered in Hα light

M. Stupar,[1,2] Q. A. Parker [1,2]

[1]*Department of Physics and Astronomy, Macquarie University, Sydney 2109, Australia*
[2]*Australian Astronomical Observatory, P.O. Box 296, Epping, NSW 1710, Australia*

**ABSTRACT**
During detailed searches for new Galactic supernova remnants (SNRs) in the Anglo Australian Observatory/United Kingdom Schmidt Telescope (AAO/UKST) Hα sur- vey of the southern Galactic plane, we also uncovered, for the first time, possible associated Hα emission in the vicinity of about 24 known Galactic SNRs previously known solely from radio or X-ray observations. The possible optical counterparts to these known SNR were detected due to the 1 arcsecond resolution and 5 Rayleigh sensitivity of this Hα survey. The newly discovered emission frequently exhibits the typical filamentary form of other optically detected SNRs although sometimes the Hα emission clouds or fragmented filaments largely inside an SNR extend over the radio border. It is true that superposition of general diffuse and extended Galactic emission in the region of these remnants is a complicating factor, but for many optical candidates the Hα emission provides an excellent morphological and positional match to the observed radio emission so that an association seems clear.

We have already published Hα images and confirmatory spectral observations for several of the best optical counterparts to known SNRs but for completeness and convenience we include them in our complete catalogue of previously known radio detected SNRs for which we have now uncovered Hα optical emission. For better visualisation of the optical emissions from these faint supernova remnants and to enhance some low surface-brightness features we also present quotient images of the Hα data divided by the accompanying broad-band short red (SR) data. Out of 274 Galactic SNRs currently catalogued and detected in the radio only ~ 20% had previous optical counterparts. We may have now increased this by a further third by adding a further 24 candidate optical counterparts.

# 1 INTRODUCTION

One of the most violent events in the universe is the explosion of supernovae. They can release energies of the order $10^{49}$ to $10^{51}$ erg. The shock-wave and ejected material expands into the surrounding ISM, sweeping up gas and dust and creating a shell. This shell of ejected and swept up material is known as a supernova remnant (SNR). SNRs have average lifetimes of 100,000 years and typical physical diameters of ~100 parsecs in their evolved state (Case & Bhattacharya 1998).

This ejected material forms the shock, causing a rise in ambient temperature and releasing high velocity particles that produce non-thermal emission. Essentially, accelerated electrons spiralling in the associated magnetic fields produce synchrotron radiation from the SNR that makes them easily detected radio sources with negative spectral index. Gamma rays are also produced as a result of proton interactions with the surrounding medium with photon energies ranging up to TeV. SNRs are often also seen in X-rays, but with a morphologically different form compared with their radio counterparts, generating both, thermal and non-thermal radiation.

Although SNRs are most conveniently studied via their radio signatures due to their strong emission in this part of the electromagnetic spectrum, historically, the first remnants were detected in the optical (examples are the Crab nebula or Tycho's supernova remnant). Systematic radio investigations and searches for SNRs began in 1950s with the work of Hanbury Brown & Hazard (1952) and Habury Brown (1954) so that today there are 274 Galactic SNRs catalogued by their radio emission. Indeed this has long been considered as a prerequisite for any true SNR identification (Green 2009) where the common classification criterion is a negative radio spectral index. This is typically expressed as $S(\nu) \propto \nu^{\alpha}$ where $S$ is the flux density and $\nu$ the given radio frequency with typical values of $\alpha \cong -0.5$ (though some-times lower values are accepted) and with $\alpha > -0.5$ being typical for younger remnants.

After identification more detailed SNR classification is based primarily on their radio morphological structure. There are three main classifications applied: shells, plerionic (Crab-like) and mixed morphology (composite). Interestingly, the first classification of shell SNRs was based on optical spectroscopy being Balmer-line dominated and oxygen-rich etc. (see classification of the Magellanic Clouds sample in Mathewson et al. 1984) though criteria used depend on signatures of shock-excitation from certain emission-line ratios such as [SII]/H$\alpha$ (e.g. Fesen, Blair & Kirshner 1985). Hence, before we can use optical spectroscopy to diagnose the expected shock excitation exhibited by SNRs, the object must first be detected in narrow-band images of H$\alpha$, [O III] or [S II]. This is so that we know where best to place the spectrograph slit as the optical SNR filamentary structures are narrow and vary in intensity. This is a main topic of this paper, revealing new optical emission likely associated with known radio remnants as a precursor to detailed spectroscopic follow-up.

When the SNRs optical emission is not obscured by intervening dust, especially at low Galactic latitudes, sensitive H$\alpha$ surveys can reveal a clear image of the SNRs optical filaments or emission structures. One such capability is provided by the high sensitivity (5 Rayleigh), decent 1 arcsecond resolution of the Anglo Australian Observatory/United Kingdom Schmidt Telescope (AAO/UKST) H$\alpha$ survey of 4000 square degrees of the southern Galactic plane. The digitised survey is available on-line[1] as the SuperCOSMOS H$\alpha$ survey (SHS hereafter) (Parker et al. 2005). SNRs, if detected in the optical, are often seen in the morphological forms of emission clouds or, more commonly, narrow, elongated filaments and arcs. These can be highly correlated with radio emission, but are mostly fragmented in nature (in the case of very old remnants, e.g. Stupar & Parker 2009) and usually located inside (occasionally outside) or on the boundaries of the observed radio remnant. Such H$\alpha$ survey discovery images are subsequently often supported with imaging of the same area using narrow-band oxygen [O III] or sulphur [S II] filters as these emission lines are also prominent in SNRs. Such imaging can reveals additional filaments and diffuse emissions (see example of [O III], [S II] and H$\alpha$ images of SNRs G126.2+1.6 and G206.9+2.3 in Rosado (1982) or in [O III] and [N II] with CCD in Boumis et al. (2007) and combined H$\alpha$ + [N II] and [S II] CCD image (of SNR HB21) in Mavromatakis, Xilouris & Boumis (2007)).

Optical spectroscopy can be a powerful, additional piece of evidence in the unequivocal recognition of an SNR, regardless of whether it has been previously known from radio (or X-ray) observations or even if it is primarily an optical remnant without prior existing non-thermal radiation being recognised. For example (see Stupar, Parker & Filipovic 2008) where H$\alpha$ optical detection and spectroscopic follow-up was used to uncover and confirm new Galactic SNRs prior to any radio detection. The ratio of key diagnostic optical emission lines, particularly [S II]/H$\alpha > 0.5$ is frequently used to confirm the shock excitation of the observed filaments as consequence of a likely supernova explosion, and therefore the existence of an SNR. This is often further corroborated with detection of other forbidden lines like [O I], [O II], [O III], [N II] or other Balmer lines (Fesen, Blair & Kirshner 1985).

Hence, optical spectral observations are used to provide the best estimates of the SNRs shock conditions, including speeds behind the shock where we have radiative spectra with Balmer and other forbidden lines. However, some young SNRs show non-radiative shocks, which appear as pure Balmer line filaments, without common forbidden lines. These are produced as a high speed shock passes through low density neutral gas. This generates a collisionless shock front where hydrogen atoms flow to the post-shock area and show (before they are ionised), faint narrow Balmer emission.

Some young remnants can also be seen in the optical, like Cassiopeia A. In these cases their morphological optical structure generally follows the radio emission, often delineating a shell structure. Some SNRs are still detected optically even when the remnant is middle-aged and some are even detected in the final, dissipation phase, where they are dissolving into the ambient ISM. Indeed, in these cases only optical detection of the remnant is often possible (see Stupar, Parker & Filipovic 2008). The observed morphological structure of an SNR as well as the question of whether the remnant can be detected in the optical (e.g. via H$\alpha$ imaging) or in the radio continuum, depends on the local ISM environment where the supernova explosion occurred and specifically the electron temperature, density and magnetic field strength of the SNR-ISM interaction. In those remnants where excellent correspondence between radio continuum emission with H$\alpha$ emission is present, it is clear that non-thermal electrons are accelerated and magnetic field intensified in close proximity to the thermal H$\alpha$ emission. However, only a very small component of H$\alpha$ emission will be seen if the mechanical energy from the supernova explosion propagates into a very low density, hot ISM. At the same time non-thermal emission will arise in different forms and instabilities will occur under the influence of magnetic fields (Cram, Green & Bock 1998). Hence, in order to see remnants in H$\alpha$, propagation of the blast wave should be into a cool and higher density ISM. This is not a strict rule though, as interstellar extinction, prevalent in the Galactic plane at low latitudes where many SNRs are found, blocks optical (H$\alpha$) radiation. Under these conditions only radio continuum detection of remnants is possible as the longer wavelengths cut through the dust.

In this work we present the results of a search for H$\alpha$ emission in the vicinity of known supernova remnants that have been previously discovered via detection of their non-thermal radio emission. Our starting point was the SHS (Parker et al. 2005), which we have already successfully employed to search for new SNRs based purely on their optical emission characteristics. Many were subsequently confirmed via optical spectroscopy and previously unrecognised radio counterparts (Stupar, Parker & Filipovic 2008). For this specific search we used the current Catalogue of Galactic supernova remnants prepared by Green (2009) and carefully selected remnants for which there were no previously reported optical detections or spectroscopy. For delineating the radio boundaries of known SNRs we predominantly used the PMN radio continuum survey at 4850 MHz (Condon, Griffith & Wright 1993), as most known Galactic SNRs have been detected at this frequency (see Stupar et al. 2005). We also used the SUMSS radio continuum surveys at 843 MHz (Bock, Large & Sadler 1999) and for declinations north of $\delta$=-40°, the NVSS survey at 1.4 GHz (Condon et al. 1998).

## 2 VISUAL INSPECTION OF THE SHS

Several steps were undertaken in performing a visual inspection of the SHS to uncover any optical emission from known Galactic SNRs. First, we inspected the 16× blocked down FITS data of every SHS survey field containing a known SNR according to the most comprehensive list of known remnants from Green (2009) and using the latest online catalogue version at the beginning of 2010. The online SHS survey provides a complete fits image of the central 25 sq. degrees of each survey field in this blocked-down form where the images have ~11 arcsec pixels. This permits entire survey fields to be downloaded for quick inspection. These survey field images are useful for detecting large-scale, low- surface brightness structures. Unfortunately, the poor resolution of this blocked down data prevented the detection of the very fine, coherent, emission filaments and structures (see Stupar & Parker 2009) that are a tell-tale optical signature of some SNRs. The data were, however, useful in the detection of more diffuse extended emission regions.

One of us (MS), also undertook direct visual inspection of the original Hα survey films at the plate library of the Royal Observatory Edinburgh where the AAO/UKST Hα survey material is archived. This involved painstaking scrutiny of all relevant Hα survey fields on a light-table. A hand-held viewer and microscope were employed during the scanning process. Candidate emission structures coinciding with the positions of known remnants were carefully noted. Of course full 0.67 arcsecond/pixel (10μm) resolution digital data can also be downloaded direct from the SHS for areas up to 30×30 arcmin in size for both the Hα survey and accompanying broad-band short-red (SR) data as every Hα image was followed with a (usually) contemporaneous continuum exposure (see details in Parker et al. 2005). The final step was thus to double check the veracity of our candidates identified from either the blocked-down fits images or from our direct visual inspection of the original survey films using the full resolution on-line pixel data. If Hα emission in the region of some known remnant was found in the form of filaments or emission clouds then both the Hα and SR full resolution data was downloaded from which quotient images (Hα divided by the matched SR counterpart) were created. These quotient images could be enhanced to reveal and de- fine the presence of fine emission structures. For the largest extended optical emissions it was necessary to mosaic several 30 arcmin regions together to ensure full coverage of the remnant. The Hα or quotient images of the candidate optical SNR counterparts were then overlaid with data from the PMN, SUMSS and NVSS (when available) radio continuum surveys. The radio data is presented as flux-level contours over the quotient or Hα grey-scale images. The radio boundaries of the remnant are also estimated and the relative positions of the optical emission with respect to the radio emission were obtained.

### 2.1 Selection criteria

Due the nature of many of the low-latitude SHS fields many of those containing radio remnants exhibit a wide range of optical emission nebulosities on various scales that are likely unrelated HII regions. This can confuse the situation when searching for bona-fide optical counterparts. Indeed we have already rejected many such possible false associations of radio remnants with Hα emission that we ascribe to unrelated HII regions and they form no part of this paper (some may later turn out to be associated pending spectroscopic observation). However, although we admit this process is somewhat subjective our selection of candidate optical remnants is done on the basis of several criteria:

Reject association if:

- No clear spatial/morphological association between the remnant and observed optical emission
- Optical emission that appears connected/related but that extends over a much wider angular area than that of the SNR
- Emission exhibits general characteristics of known HII regions e.g. presence of dust lanes, internal cavities with stars, more diffuse, cloudy regions etc.

Consider a likely association if:
- Optical emission is well correlated spatially and morphologically with SNR radio data
- Optical emission has the typical form of other known optically detected remnants such as fine filaments and structures that follow the radio contours
- Any available optical spectra confirm the presence of shocks, a typical diagnostic of optically detected remnants.

## 3 THE NEW CATALOGUE OF GALACTIC SNRS UNCOVERED IN Hα LIGHT

Table 1 lists the 24 known Galactic SNRs (mostly from radio observations) where we report the detection of likely associated optical Hα emission for the first time. Columns 1 to 3 are taken from Green (2009) and classified following the normal

Galactic coordinate convention. If only one or two sources of optical emission were detected, their size in arminutes is given in column 4, and if there is more than one, it is stated in the last column. Usually, optical SNRs are detected in three morphological classes: nebulous emission clouds, a series of fragmented optical filaments or as a mixture of both. This is indicated in columns 5 and 6.

Note that for three of the most prominent, new, optically detected remnants listed in Table 1 we have already undertaken and published a comprehensive multi-wavelength and spectroscopic study but they are included here for completeness. These are for G315.1+2.7 (Stupar, Parker & Filipovic 2007a), G332.5-5.6 (Stupar et al. 2007b) and G279.0+1.1 (Stupar & Parker 2009). An overall introduction to this work is also given in Stupar, Parker & Filipovic (2007c). In this section we provide brief notes on all the candidate

**G4.2-3.5.** This remnant is shown on Fig. 1 and is an excellent example of pure SNR Hα filaments without more dif- fuse cloud emission components. The optical filaments provide an excellent match on the N-E and N-W sides of the radio shell emission at 1.4 GHz from the NVSS radio continuum survey. The most prominent group of filaments are enclosed within the radio shell and have a size of 9×14 arcminutes. Our preliminary follow-up spectral observations of the brightest N-E filament gave a diagnostic emission line ratio of    [S II] / Hα = 0.65 which is typical for an SNR shock and fits well within the SNR regime as described by Fesen, Blair & Kirshner (1985). Other typical SNR emission lines such as very strong optical counterparts to the known radio SNRs.

[O II] at 3727 Å and strong [O III] at 5007Å were also noted (Stupar & Parker in preparation). This is one of our best examples of a new, clear optical counterpart to the known radio remnant.

**G5.5+0.3.** Brogan et al. (2006) reported this SNR as a shell 12×15 arcmin across after radio observations at three frequencies of 330, 1500 and 2720 MHz. Fig. 2 presents the SHS Hα image of this area overlaid with PMN 4850 MHz radio contours. A similarity in the form of SNR G5.5+0.3 between the PMN radio image and the one given in Brogan et al. (2006) at 330 MHz is noted, while the NVSS radio map more or less follows   the optical emission. For this SNR the match of the radio and optical data is confined to a small part of a large ~1.5˚ Hα arc, with the   arc opened-up towards the east.

From the point of view of optical morphology, we could say that this arc defines the extent of the optical SNR, but in fact the radio observations imply that just a part of the optical structure is likely associated with the SNR. This could be also connected with the Brogan et al. (2006) classification of SNR G5.5+0.3 as their class II - where they only state they are "fairly confident" that they have detected an SNR and that the source is possibly confused with more general thermal emission. However, indicative Hα filaments are clearly seen in the north and south of G5.5+0.3 (also covered with radio emission). Deeper optical imagery and   follow-up spectroscopy is needed to solve the problem of the true nature of G5.5+0.3 as well as a detailed spectral analysis of the whole 1.5˚ Hα arc to determine which parts (if any) are actually associated with the radio remnant.

Table 1. List of known Galactic SNRs seen in optical light for the first time. Columns 1 to 3 are from Green (2009)

| Standard ID | R.A. J2000.0 | δ J2000.0 | Optical extent (arcmin) | Emission cloud | Filament(s) | SHS field # | Remarks |
|---|---|---|---|---|---|---|---|
| G4.2-3.5 | 18 08 55 | -27 03 | 19× 14 | N | Y | 711 | Chain of filaments |
| G5.5+0.3 | 17 57 04 | -24 00 | – | N | Y | 794 | Arc   2˚ |
| G6.1+0.5 | 17 57 29 | -23 25 | 3× 11 | Y | Y | 794 | Prominent arc |
| G6.5-0.4 | 18 02 11 | -23 34 | 5× 8 | Y | Y | 794 | |
| G9.9-0.8 | 18 10 41 | -20 43 | 5× 4.5 | Y | Y | 880 | |
| G11.1-1.0 | 18 14 03 | -19 46 | 13× 9 | N | Y | 880 | |
| G15.1-1.6 | 18 24 00 | -16 34 | 25× 25 | Y | Y | 970 | Three larger H  α emissions |
| G16.8-1.1 | 18 25 20 | -14 46 | 8× 16 | Y | Y | 970 | |
| G18.9-1.1 | 18 29 50 | -12 58 | 35× 30 | Y | Y | 1060 | Mostly filaments |
| G32.8-0.1 | 18 51 25 | -00 08 | 17× 10 | Y | Y | 1332 | Few filament groups |
| G279.0+1.1 | 09 57 40 | -53 15 | – | Y | Y | 275 | 14 filaments/clouds    [1] |
| G286.5-1.2 | 10 35 40 | -59 42 | 3× 3 | N | Y | 175 | |
| G291.0-0.1 | 11 11 54 | -60 38 | – | N | Y | 176 | Two groups of filaments |
| G315.1+2.7 | 14 24 30 | -57 50 | 0.8× 11 | N | Y | 231 | Large arcuate filament    [2] |
| G315.4-0.3 | 14 35 55 | -60 36 | 1  × 3;1× 4 | N | Y | 182 | Two arcuate filaments |
| G317.3-0.2 | 14 49 40 | -59 46 | 10× 10 | N | Y | 183 | |
| G320.6-1.6 | 15 17 50 | -59 16 | – | Y | N | 184 | Irregular cloud |
| G332.4+0.1 | 16 15 20 | -50 42 | 10× 2 | N | Y | 289 | Kes 32 |
| G332.5-5.6 | 16 43 20 | -54 30 | – | Y | Y | 236 | Few filaments   [3] |
| G336.7+0.5 | 16 32 11 | -47 19 | 3× 3 | Y | Y | 349 | |
| G340.4+0.4 | 16 46 31 | -44 39 | 0.2× 3.5 | N | Y | 414 | |
| G350.0-2.0 | 17 27 50 | -38 32 | – | N | Y | 484 | clouds/filaments |
| G359.0-0.9 | 17 46 50 | -30 16 | – | N | Y | 631 | Few small nebulosities |
| G359.1-0.5 | 17 45 30 | -29 57 | 9.5× 8.5 | Y | Y | 631 | |

[1] See MNRAS (2009) 394, 1791;    [2] See MNRAS (2007) 374, 1441;    [3] See MNRAS (2007) 381, 377

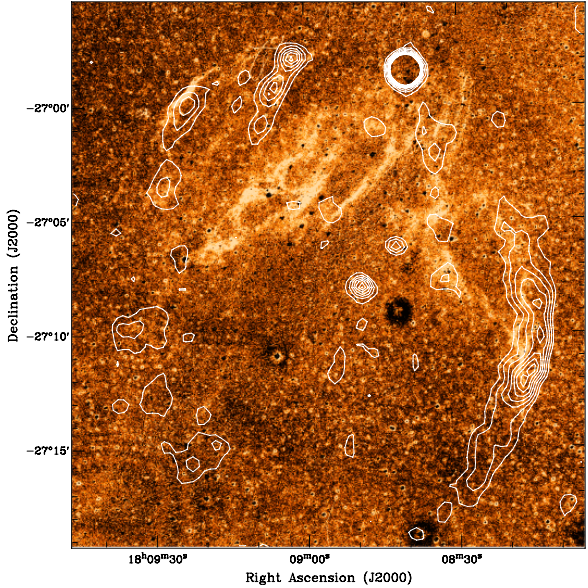

**Figure 1.** Quotient optical image (Hα divided by SR) of known Galactic SNR G4.2-3.5 overlaid with NVSS 1.4 GHz contours from 0.001 to 0.01 Jy beam⁻¹. There is an excellent match of the radio and Hα emission for this remnant to the N-E and N-W sides of the radio shell making this one of our clearest and best optical counterparts. Also note the characteristic series of optical filaments.

**G6.1+0.5.** This Galactic remnant (Fig 3) was discovered by Brogan et al. (2006) and observed in the radio at 330, 1500 and 2720 MHz. Two identical radio spectral in- dices of α = −0.9 were obtained between these three given radio frequencies which is completely typical for Galactic SNRs. The PMN (green) radio contours cover most of the optical emission, especially on the east side, where the set of fine, narrow, optical filaments is evident and crosses the PMN 4850 MHz contours. We compared the Brogan et al. (2006) 330 MHz image with the SHS Hα image and noted the general similarity in extent. However, the 6 arcmin long, narrow, optical filament, clearly dominant in Hα towards the centre of the image, most probably represents a shock on the western side of G6.1+0.5. This optical filament is also clearly identified in the NVSS 1.4 GHz image which is shown in expanded form to the upper right corner on Fig 3 and with more details in Fig. 4, where the red colour contours show the flux at this frequency. No significant radio flux was noted to the east side of G6.1+0.5 in the 1.4 GHz NVSS image.

We consider that this 6 arcmin narrow, optical fila- ment (Fig. 4), together with its excellent match at 1.4 GHz, while clearly associated could even originate from a different radio SNR shell as it is well known that remnants ex- hibit different morphological structures at different radio frequencies. It is clear that this area needs further radio fol- low-up at higher resolution together with optical spectroscopy of the prominent 6 arcmin filament, the diffuse emission cloud to the east and the fine filaments putatively associated with G6.1+0.5 and clearly seen in Fig 3 to the North.

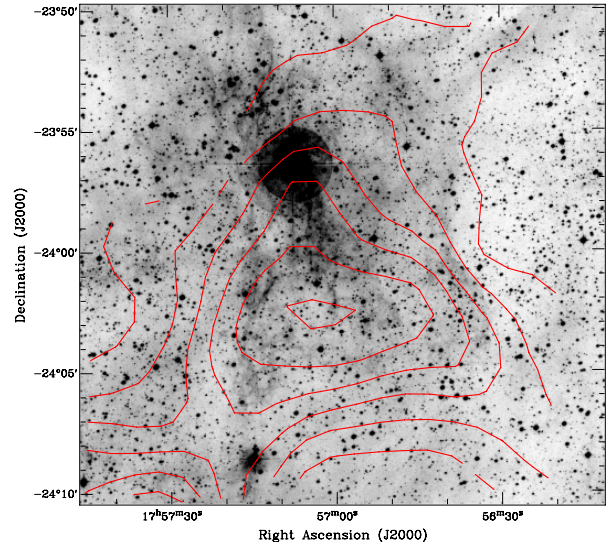

**Figure 2.** A 20 × 20 arcminute Hα grey-scale image of SNR G5.5+0.3 showing only the area clearly seen in Hα though the optical emission appears to extend for ∼ 1.5˚. It is not clear how much of this emission is directly associated with the 12×15 arcmin radio remnant. The Hα image is overlaid with PMN radio contours from 0.02 to 0.3 Jy beam⁻¹. When comparing the PMN and NVSS fluxes for this rem- nant it was noted that the PMN flux (shown) is similar to the Brogan et al. (2006) 330 MHz radio image of the remnant while the NVSS data more or less follows the narrow central part of the Hα morpho- logical structure.

**G6.5-0.4.** This SNR forms a distinct radio shell about 18 arc- min in diameter and is situated on the south-east side of the known mixed morphology SNR W28 (G006.4- 00.1). The SHS image (Fig. 5) shows low and medium level Hα emis- sion in the central and south-eastern part of the remnant while on the south west there is a very strong condensation of opti- cal nebulosity (approx. position RA=18ʰ02ᵐ00ˢ and δ=−23˚42′(J2000.0)). This compact emission has previously been classified as a suspected HII region BFS 1 in Blitz, Fich & Stark (1982) but without specific measurements being pre- sented in their catalogue of CO radial velocities toward Ga- lactic HII regions. Interestingly, in this area of strong Hα emission, SIMBAD notes the X-ray source ROSAT 1RXS J180158.8-234134, and the infrared source IRAS 17588-2340 as well as the point like PMN radio source marked as PMN J1801-2339. We propose that these optical emissions clearly seen in Fig. 5 are, in fact, the optical counterparts to the SNR but we need optical spectra to confirm their nature. Our literature search for optical spectra (e.g. to confirm/refute the HII nature of filaments) failed. Note it is common that SNR optical emission is sometimes located close to but outside the observed radio borders. Any possible connection between the prominent HII region W28 and G6.5-0.4 could be uncovered with detailed optical spectroscopy. Note that at the lowest radio flux levels to the north, east and south the contours are not completely closed because of missing radio flux in  this area due to  blank fields most probably the result of solar interference during the PMN observations. If this effect is taken into account, then it is possible that the Hα filaments in the south-east do have a radio counterpart. Furthermore, there is one more prominent Hα emission region centred at RA=18h02m33s and δ=−23°39′(J2000.0) which follows the general shape of the PMN radio contours but also extends further to the south west. The equivalent NVSS image of G6.5-0.4 does not show any shell structure only a few, small, disconnected concentrations of radio emission including a localised detection to the south which has a close positional coincidence with the same optically suspected HII region at RA=18$^h$02$^m$00$^s$ and δ=−23°42′(J2000.0).

red) from 0.07 to 1.7 Jy beam$^{-1}$ and NVSS 1.4 GHz contours (in white) from 0.001 to 0.04 Jy beam$^{-1}$. The PMN contours present a smoothed circular region of radio emission due to the ~5 arcmin survey resolution, while the higher resolution NVSS data shows a clear, but fractured radio shell. There is an excellent match between the NVSS 1.4 GHz flux and the Hα emission around the shell but with the strongest Hα emission corresponding both to the peak in the PMN data and in the NVSS data indicating a very clear connection. Brogan et al. (2006) found the average negative radio spectral index for this object to be α = −0.6 which is typical for such shell structured remnants, and deemed sufficiently conclusive for it to be identified as a bona-fide SNR. However, they also noted possible confusion with thermal emission and the need for future radio observations at higher resolution and sensitivity. We have now uncovered clear optical emission that follows the shell outline and matches the strongest radio emission with matching Hα prominence indicating that they must be connected and that the optical counterparts are real. Follow-up optical spectroscopy should further clarify the situation and indicate if there is a thermal emission component.

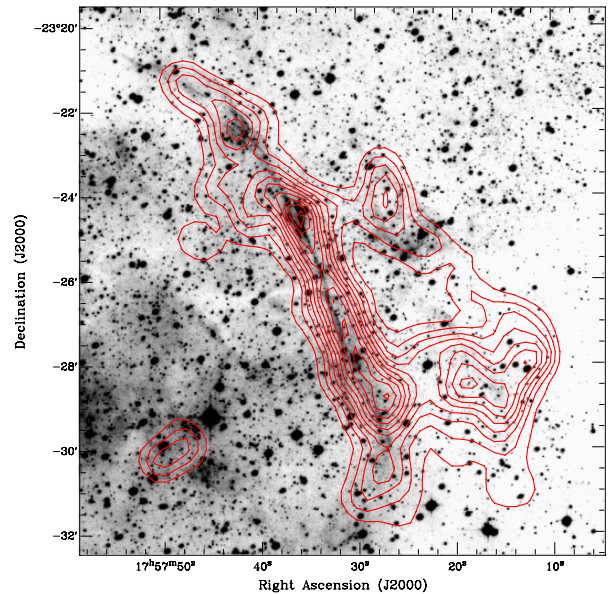

**Figure 4.** G6.1+0.5 as seen in the NVSS survey at 1.4 GHz with red flux contours from 0.006 to 0.035 Jy beam$^{-1}$ overlaying the SHS optical image. No flux is noticed east of the 6 arcmin long filament, but on the right side more radio emission is present. The NVSS detection seems morphologically distinct from the coarser resolution PMN data.

**G11.1-1.0.** This known Galactic SNR appears in Hα light in the form of two main parallel emission arcs separated by about 6 arcmin east-west and with a southern hook on each arc. They are about ~8 arcmin in length and are clearly part of the same structure given the strong morphological similarity. Such structures are also typical of optical remnants. Both arcs present a higher intensity of Hα emission on the southern side. The eastern arc is also stronger and has an excellent

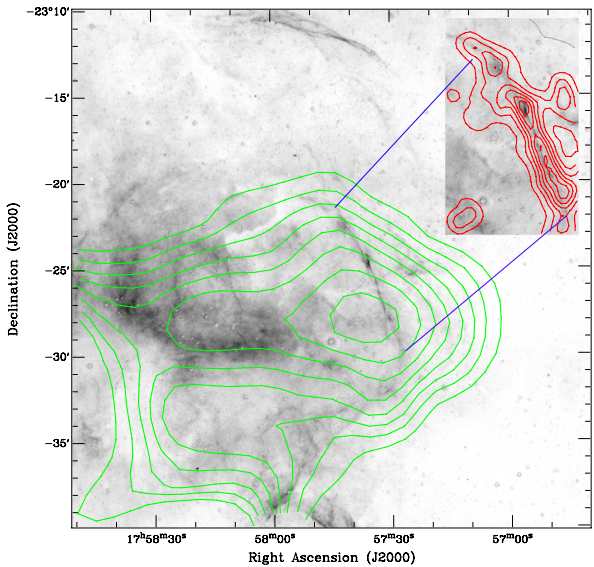

**Figure 3.** Galactic SNR G6.1+0.5 showing the SHS quotient Hα/SR image overlaid with (green) PMN 0.08 do 0.5 Jy beam$^{-1}$ contours. A strong similarity between the PMN and 330 MHz radio data was noted for this remnant (Brogan et al. 2006). However, the NVSS 1.4 GHz image of G6.1+0.5 is somewhat different though the prominent 6 arcminute Hα arc seen towards the centre of the image is also clearly defined at 1.4 GHz. This is shown in expanded form in the upper right corner where the excellent Hα and NVSS match is verified. The NVSS red contours go from 0.004 to 0.03 Jy beam$^{-1}$.

**G9.9-0.8.** The SNR G9.9-0.8 has a typical radio shell morphological structure 12 arcmin in diameter at 330 MHz as reported in Brogan et al. (2006). Fig. 6 shows the SHS Hα data as a grey scale image overlaid with PMN contours (in match with the NVSS radio emission at 1.4 GHz (see Fig. 7 with green contours from 0.002 to 0.03 Jy beam⁻¹), which marks the eastern boundary of the SNR at this frequency. The SNR is also detected in the PMN survey at 4850 MHz shown by red contours from 0.03 to 0.9 Jy beam⁻¹ in Fig. 7. The PMN data is effectively centered on the optical emission but due to the low resolution of this survey the radio contours do not follow the optical and NVSS arc curvature though the PMN peak flux does match the location of the strong eastern arc. Also, a very prominent 2 arcmin filament can be seen in the Hα image positioned to the south-east but without any particular radio emission.

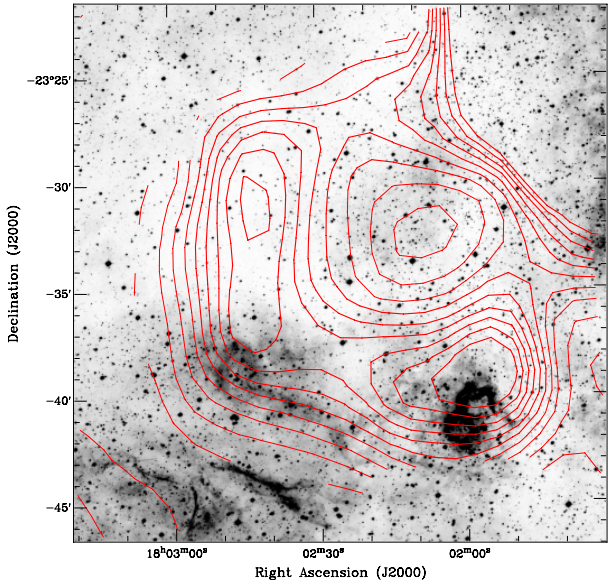

**Figure 5.** Hα image of SNR G6.5-0.4, situated east of the known mixed morphology SNR W28 (G006.4-00.1), overlaid with PMN contours from 0.01 to 1 Jy beam⁻¹. Hα emission can be noticed in the central and eastern part of the remnant together with a group of more prominent filaments to the south-east which run parallel to the radio contours and also extend beyond the apparent radio borders. Follow-up optical spectra should confirm the true nature of these filaments but we propose that they are the optical counterparts. The approximately circular, condensed optical emission at RA=18ʰ02ᵐ00ˢ and δ=−23°42′(J2000.0) is a likely overlapping but unrelated HII region BFS1 (Blitz, Fich & Stark 1982) which also appears to be detected in the PMN, and in the ROSAT and IRAF as X-ray and infrared sources. High resolution optical spectra could reveal the true nature of this strong Hα emission. W28 is situated off to the west and north-west and is not shown in the image.

Note that apart from the structured filamentary Hα emission of Fig. 7, there are also areas of dust extinction and a general wash of almost uniform background emission to the east, partially in and out of the SNR border. We believe this optical emission from known Galactic SNRs is probably unconnected with the remnant and is instead part of the general complicated ISM in this direction. However, the morphologically distinct arcuate filaments that are typical in optical SNRs are clearly connected with the radio data making a very strong positional and morphological link with the known SNR. Planned follow-up optical spectroscopy should confirm this.

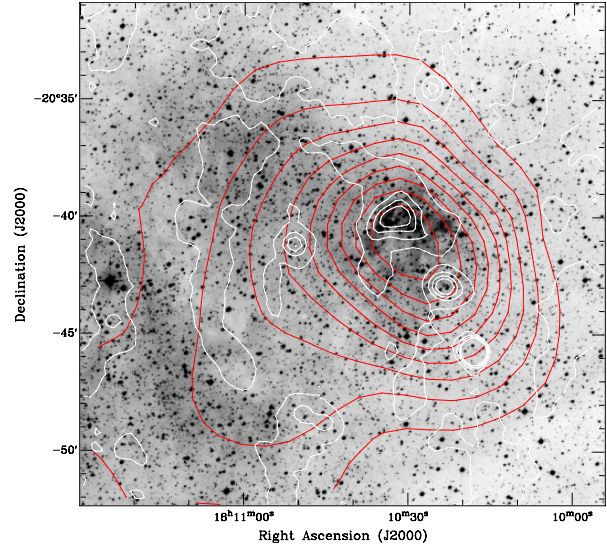

**Figure 6.** Hα image of SNR G9.9-0.8, overlaid with PMN contours (in red) from 0.07 to 1.7 Jy beam⁻¹ and NVSS 1.4 GHz contours (in white) from 0.001 to 0.04 Jy beam⁻¹. The PMN contours present a smoothed circular region of radio emission due to the ~5 arcmin survey resolution, while the higher resolution NVSS data shows a clear, but fractured radio shell. There is an excellent match between the NVSS 1.4 GHz flux and the Hα emission around the shell. The strongest Hα emission corresponds both to the peak in the PMN data and that in the NVSS data indicating a very clear connection.

**G15.1-1.6.** We first published an Hα image of Galactic SNR G15.1-1.6 in the AAO Newsletter, (Stupar, Parker & Filipovic 2007c) where we first discovered and proposed the identification of an optical Hα emission counterpart to this radio SNR. Fig. 8 shows the Hα image overlaid with PMN radio emission where the green contours represent flux at 4850 MHz between 0.08 and 0.37 Jy beam⁻¹. The Hα data has an excellent positional match with the radio emission over the whole area of the remnant. Recently, Boumis et al. (2008), independently uncovered optical emission from this remnant (but post-dating our reporting of this discovery in the AAO Newsletter) and observed this remnant spectroscopically. They confirmed the shock emission typically revealed by SNR optical counterparts and also reported possible contamination of the remnant with a HII region in the eastern area. They also noticed strong [O III] emission in 5007Å. We note that our SHS Hα imaging is superior to Boumis et al. (2008) with a clear Hα shell being evident with a particular concentration to the eastern side. The clear optical SNR counterpart is confirmed.

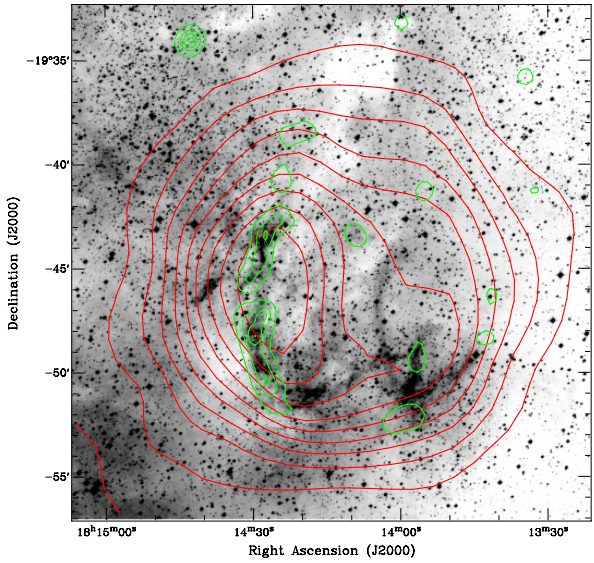

**Figure 7.** Greyscale SHS Hα image of known Galactic SNR G11.1-1.0 overlaid with PMN 4850 MHz contours (red) from 0.03 to 0.9 Jy beam$^{-1}$ and NVSS 1.4 GHz contours (green) from 0.002 to 0.03 Jy beam$^{-1}$. Note the NVSS flux contours completely follow the eastern arc structure in Hα while the PMN contours are effectively centred on the Hα structures indicating a clear link between the optical and radio data.

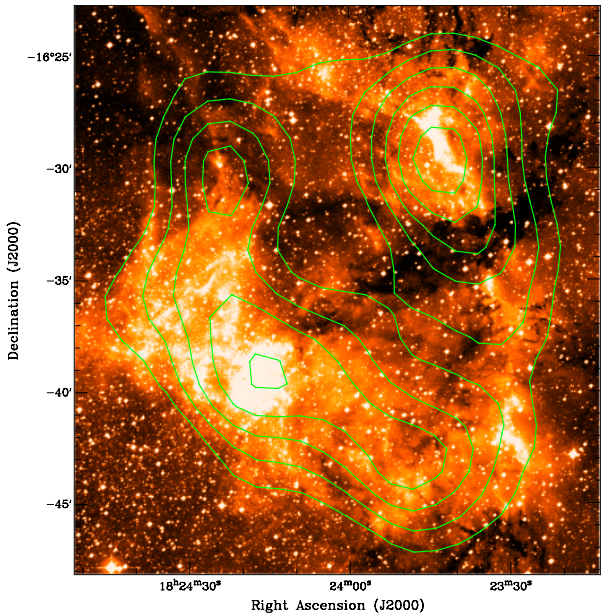

**Figure 8.** Galactic SNR G15.1-1.6 in Hα light as first published in Stupar, Parker & Filipovic (2007c) and subsequently reported by Boumis et al. (2008). The image is overlaid with the PMN radio emission map where green contours represent flux at 4850 MHz between 0.08 and 0.37 Jy beam$^{-1}$. It is clear that the optical emission is of SNR origin and strongly correlated with the radio data forming almost a complete optical shell.

G16.8-1.1. We believe the provenance of this SNR is in doubt although radio observations (Reich et al. 1986) confirm non-thermal radiation. Fig. 9 shows the SHS Hα image in the vicinity of SNR G16.8-1.1 overlaid with PMN (red) radio contours from 0.08 to 1.14 Jy beam$^{-1}$ and NVSS (green) contours from 0.001 to 0.34 Jy beam$^{-1}$. It is clear the PMN and NVSS data peaks coincide with the strong Hα emission which, however, has the typical form of a HII region and not an SNR. As can be expected given the closeness in frequency, we found overall agreement in the radio morphological structure of G16.8-1.1 between the PMN and the Reich et al. (1986) 4750 MHz observations which, however, have better resolution than the PMN. The NVSS 1.4 GHz data reveals a somewhat different structure to that seen in the lower resolution PMN radio map with radio flux concentrated only in the central area together with some fragmented spots in the north that match the Hα emission.

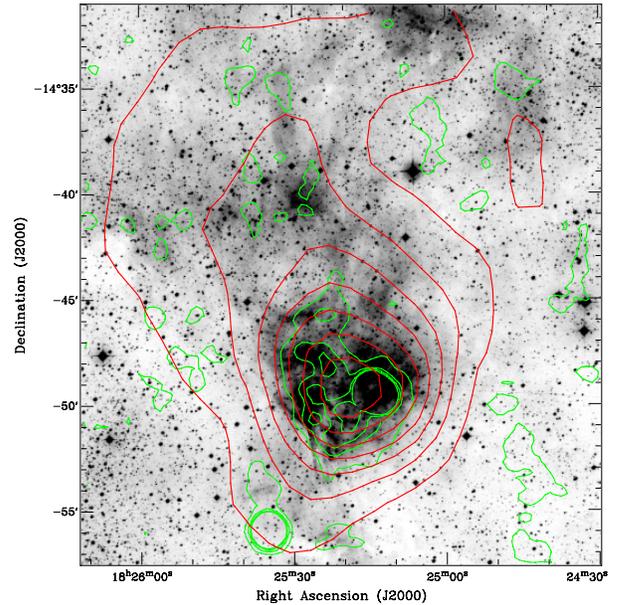

**Figure 9.** The SHS Hα image in the vicinity of SNR G16.8-1.1 overlaid with PMN (red) radio contours from 0.08 to 1.14 Jy beam$^{-1}$ and NVSS 1.4 GHz (green) contours from 0.001 to 0.34 Jy beam$^{-1}$. It is clear the PMN and NVSS data peaks coincide with the strong Hα emission which, however, has the typical form of a HII region and not an SNR. Optical spectroscopy of the various optical emission components that match the radio maps should indicate where any thermal and non-thermal emission components are located and perhaps clearly define the nature of the optical emission as originating from a HII region, SNR or a indeed mixture of both.

This is not the first time that this object has been seen in the optical and has been previously identified as HII region S50 and in the radio as RCW 164 (Rodgers, Campbell & Whiteoak 1960). Reich et al. (1986) concluded that G16.8-1.1 is, in fact, an SNR based on radio data. However, it is not clear which, if any optical components can be associated with the putative SNR. In fact Reich et al. (1986) concluded that this object is a mixture of thermal and non-thermal emission and that the radio spectral index derivation is uncertain. On the basis of their radio polarization observations, they concluded that the obvious HII region is in the front of a non-thermal source, most probably an SNR. Optical follow-up spectral observations should show the presence of thermal and non-thermal emission mechanisms and there- fore potentially the position of any optical counterpart to G16.8-1.1 if it is in fact a bona-fide SNR. One should bear in mind that the Reich et al. (1986) observations were under- taken with the Effelsberg 100m and Nobeyama 45m dishes at 4 frequencies and with a relatively "coarse" Half-Power Beam Width (HPBW) of between 2.4 and 9.3 arcmin

**G18.9-1.1.** It would be very interesting to acquire optical spectra of the optical filaments possibly associated with this remnant as the existing radio observations (Odegard 1986) exhibit different values of the radio spectral index (from -0.20 to -0.35 between 57.5 MHz and 4.75 GHz) over different parts of the remnant. It is thought that SNR G18.9- 1.1 is powered by a pulsar (Odegard 1986) but this idea was later abandoned (Fuerst, Reich & Aschenbach 1997) and then re-examined in Harrus et al. (2004). The SHS quotient image of the area given in Fig. 10 shows optical emission on the north, east and south-west sides in the form of long filaments. To the south and north-east are some additional indistinct filaments outside the radio borders which may also be related. Future spectral observations of these features could show any possible connection with G18.9-1.1.

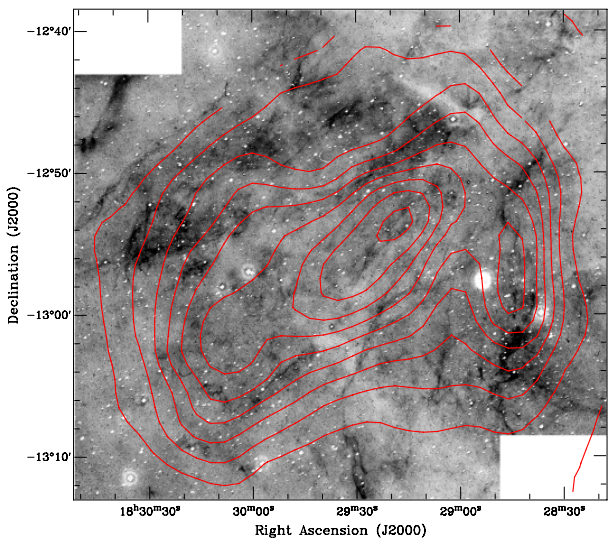

**Figure 10.** The 35×35 arcmin mosaic SHS quotient (Hα over SR) image of SNR G18.9-1.1 overlaid with PMN contours from 0.05 to 0.97 Jy beam⁻¹. Faint optical emission can be noticed to the east, north and west in the form of light, diffuse nebulosities that follow the radio contours.

**G32.8-0.1 - Kes 78.** Galactic SNR G32.8-0.1 was first associated with optical emission from the SHS in Stupar (2007) and represents one of our best examples of a very clear optical shell-like counterpart to a known radio SNR. It was independently reported by Boumis et al. (2009) before we could undertake our own detailed follow-up. They also undertook follow-up optical spectroscopy and found a very strong ratio of [S II] / Hα = 1.5 confirming shock excitation from the SNR but with a very slow velocities due to the lack of [O III] in their spectra. Fig. 11 shows the 20 × 25 arcmin SHS optical quotient image of this remnant overlaid with NVSS 1.4 GHz radio contours from 0.002 to 0.1 Jy beam⁻¹. There is an excellent match between the radio and optical emission. Indeed, the optical data reveals an elongated, oval shell far more morphologically distinct than that seen in the radio map. In the PMN survey this part of the sky was scanned with the Green Bank 30-m antenna and, although Kes 78 was detected, the definition of remnant in these data is not completely clear due to the low resolution.

**G279.0+1.1.** We have already presented a detailed study of this SNR in Stupar & Parker (2009) and it is included here for completeness. This SNR has numerous small optical filaments which extend over a ~2° area. For further information on spectra and images we refer the reader to our paper.

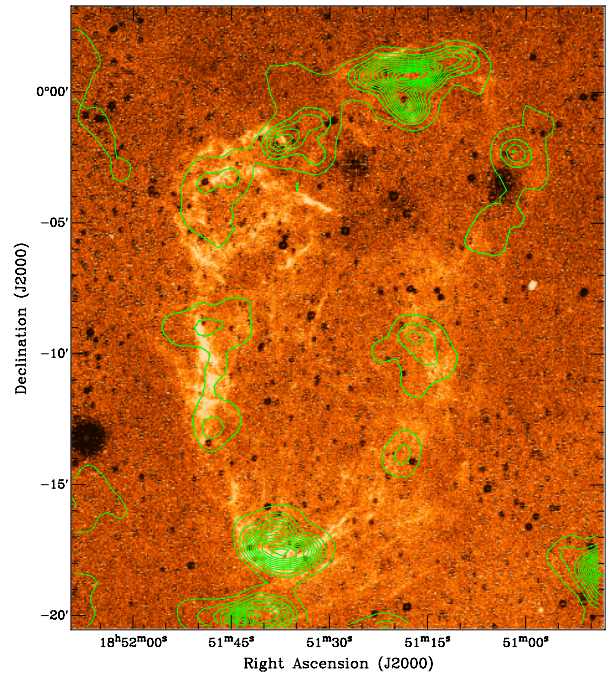

**Figure 11.** SHS quotient image of SNR G32.8-0.1 (Kes 78). There is an excellent match of the optical data with the NVSS radio contours from 0.002 to 0.1 Jy beam⁻¹ which reveals an elongated oval shell far more morphologically distinct than that seen in the radio map. Interestingly there is a group of optical filaments at the approximate position RA=18ʰ51ᵐ50ˢ and δ=−0˚11′(J2000.0) of a strong OH 1720 MHz methanol maser (see Koralesky et al. 1979).

**G286.5-1.2.** This object was first reported as an SNR in the MOST2 catalogue of SNRs at 843 MHz (Whiteoak & Green 1996) as two, parallel arc shaped filaments. The PMN survey data is of lower resolution and only shows this area as a single, extended curved structure of similar size rather than the two distinct filaments revealed by the MOST data. The first filament is some 15 arcmin in length while the second filament is situated to the north-west of this main filament though they appear to merge to the north-east. Here, there is an extension of 843 MHz emission, irregular in form and some 10 arcmin in length where an excellent match to our Hα emission (see Fig. 12 which also includes SUMSS[2] data) is seen. An excellent match between the radio and Hα emission components also exists at the southern part of the longer filament. Diffuse Hα emission to the south-east may be unrelated. Preliminary spectroscopic follow-up confirmed the shock excitation mechanism for these optical filaments and their close morphological and positional co-incidence with the known radio structure (Stupar, Parker & Filipovic in preparation). Taken together the combined spectroscopic and morphological evidence indicates that we have almost certainly uncovered the optical counterpart of this remnant.

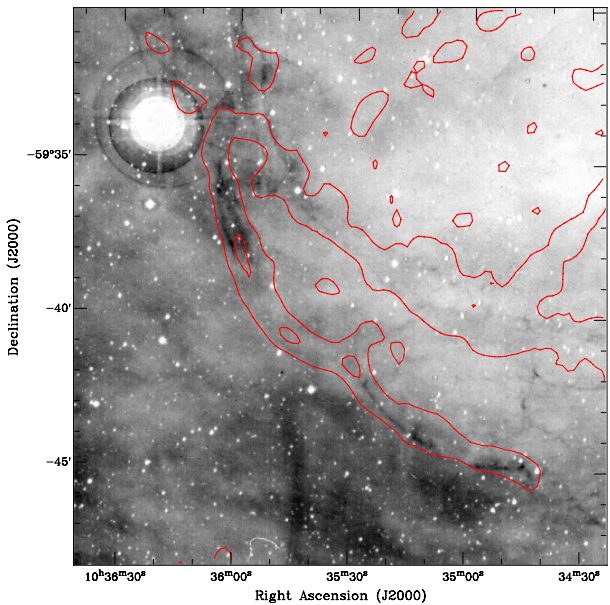

**Figure 12.** SHS quotient image of G286.5-1.2 overlaid with radio data from the SUMSS 843 MHz survey with contours at 0.008 to 0.06 Jy beam$^{-1}$ that reveal two filamentary arc-like radio structures. We used just two extreme contours of the SUMSS data of G286.5-1.2 to clearly show the excellent match between the optical and SUMSS 843 MHz radio filaments.

---

[2] SUMSS (Bock, Large & Sadler 1999) is a radio survey of the southern sky from δ=−30' to the south while (Whiteoak & Green 1996) present a catalogue of SNRs prepared with the same (Molonglo) radio telescope as the SUMSS and at the same frequency of 843 MHz.

**G291.0-0.1** or **MSH 11-62.** This SNR has a distinct, elongated radio bar about 10 arcmin in extent running al- most directly north-east to south-west and is surrounded by a more circular radio envelope with a secondary peak about 4 arcmin to the south east. The possibly associated optical emission forms an broad arc from the west to the south- east where it runs into more general diffuse emission to the far south east that may be unrelated. There are also two more intense filamentary Hα emission regions, one to the east of the main radio bar and the second just to the south of the secondary radio peak. Fig. 13 shows a greyscale SHS Hα image overlaid with SUMSS 843 MHz radio contours from 0.06 to 0.4 Jy beam$^{-1}$ where the bright, elongated central core and secondary component are seen. There is also a much weaker third peak at a similar distance to the north- northwest, all within an overall approximately circular radio envelope 15 armin across at maximum extent. The two localised areas with prominent Hα filaments that appear to be connected with the radio emission are expanded as separate regions on the right side of the Fig. 13. It can be seen that the optical filaments are concentrated on the enhanced radio rims at the edges of the remnant. There is no obvious optical emission from the central, radio bright component however. The more diffuse optical structure to the south-east extends outside the radio borders but may be unrelated. We intend to undertake follow-up optical spectroscopy to confirm the shock excited nature of the two most conspicuous filament groups, which we propose are an optical manifestation of the SNR.

**G315.1+2.7.** We have already published Hα imagery and optical spectra of a major associated optical filament of G315.1+2.7 confirming an SNR origin and including radio observations at different frequencies and we include this object here for completeness. In Hα light this SNR appears as an arcuate filament some 9 arcmin in extension. We refer the reader to our paper Stupar, Parker & Filipovic (2007a) for further details on this remnant.

**G315.4-0.3.** This SNR has been known for some time. Caswell, Milne & Wellington (1981) compared their observations of this SNR at 1415 MHz with previous radio maps at 408 and 5000 MHz of (Caswell et al. 1975) and confirmed a radio spectral index of α = - 0.47. The SNR forms an oval shell like structure about 17 arcmin across at maximum extent. Fig. 14 gives the SHS Hα image overlaid with SUMSS 843 MHZ radio contours from 0.004 to 0.03 Jy beam−1. The SHS reveals at least three regions with enhanced Hα emission interior to the remnant. The two most prominent are shown enlarged in the upper panels of Fig. 14. The eastern nebula structure at R.A.=14$^h$ 36$^m$ 46$^s$ and δ = −60°30'45'' (J2000.0) is more filamentary, while the western emission at R.A.=14$^h$ 35$^m$ 50$^s$ and δ = −60° 33' 40'' (J2000.0) is more diffuse and ~4 arcmin in size at maximum extent. On the north-east side there is an unresolved source with a steep spectral index, which is most probably extragalactic (Whiteoak & Green 1996). On the south western side of this remnant there is a compact radio source which Whiteoak & Green (1996) noted as a HII region. The HII region was also catalogued by Caswell & Haynes (1987) as GAL 315.31-00.27 based on 5

GHz hydrogen recombination line observations where they simply commented the object as a distant HII region superimposed on an SNR. However, this compact source is not registered in our Hα image, most probably due to high extinction in this area (although Schlegel maps (Schlegel, Finkbeiner & Davis 1998) showed unreliable extinction estimates for this |*b*|).

The overall conclusion from Caswell et al. (1975) and Caswell, Milne & Wellington (1981) is that much of the remnant is non-thermal in its radio emission. Besides, Caswell, Milne & Wellington (1981) could find no trace of optical emission at the position of the G315.4-0.3 radio shell in the optical southern sky surveys available at that time. However, the highly sensitivity SHS has now revealed plausible optical counterparts for the first time that do seem to delineate the inner regions of the radio shell quite nicely. Follow-up optical spectroscopy is now needed to provide the emission line diagnostics to confirm this probable association.

**G317.3-0.2.** The radio image of this remnant at 843 MHz (11×11 arcmin in size) shows two opposing arcs of radio emission (see Whiteoak & Green 1996), while the PMN data at 4850 MHz shows a more classical shell (contours on Fig. 15) but of a different size compared with the 843 MHz data, i.e. ~11×20 arcmin. The difference in morphological form seen at different radio frequencies could simply be due to the lower resolution of the PMN survey though the peak radio flux position is also different; at 843 MHz it is located in the south-east part at 0.2 Jy beam⁻¹, while in the PMN it is in the central area of the remnant with a peak of 1.27 Jy beam⁻¹.

The SHS Hα image of G317.3-0.2 (Fig. 15) reveals emission that broadly matches the PMN emission across the whole 10 arcmin central area of the remnant except to the south-east of the shell in 843 MHz (see Whiteoak & Green 1996). The increased Hα emission in the central and south-western areas has an excellent match with the PMN flux at 4850 MHz where the fractured and irregular Hα emission partially goes over the lowest flux contours of the PMN data. Spectroscopic follow-up is planned of some of the fractured Hα emission components to check for shock excitation.

**G320.6-1.6.** The SUMSS radio data of this SNR reveals two opposing almost parallel arcs of emission in an open shell-like structure where the western arc has an intense southern counterpart. Georgelin & Georgelin (1970) in their optical review of HII regions partially covered the area of G320.6-1.6 but, due to their low resolution images (~ 1 arcmin) the nature of this region was not obvious and was possibly mixed with structure from the nearby SNR G320.4-1. s (RCW 89). Our SHS quotient image shows a region of very complex morphological structures combining both emission regions and dust (see Fig. 16) where there appear to be a mixture of features associated with various HII regions and some Hα filaments which might also be SNR in origin. Some of these optical filaments are highlighted in the right panel of Fig. 16 which shows a chain of Hα filaments some 15 arcmin in length. We also compared the SUMSS 843 MHz data with PMN 4850 MHz emission and found similarity in overall structure except that the eastern side of the remnant shows two parallel arcs in the SUMSS data.

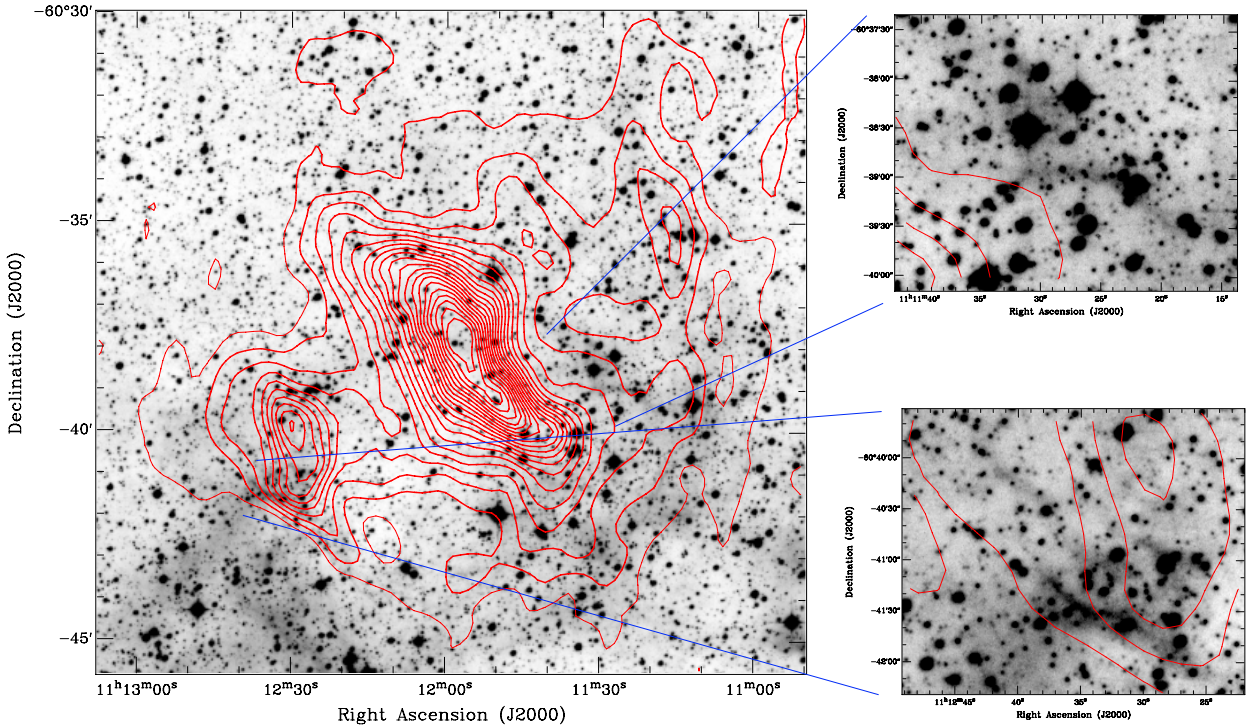

**Figure 13.** G291.0-0.1 or MSH 11-62 in Hα light with two enlarged images showing two groups of optical filaments in relation to the radio emission. The SUMSS 843 MHz contours go from 0.06 to 0.4 Jy beam⁻¹ and clearly delineate the prominent elongated central core and the weaker secondary and tertiary radio flux peaks to the south-southeast and north-northwest of the central region within the overall radio envelope which is about 15 arcmin across at maximum extent.

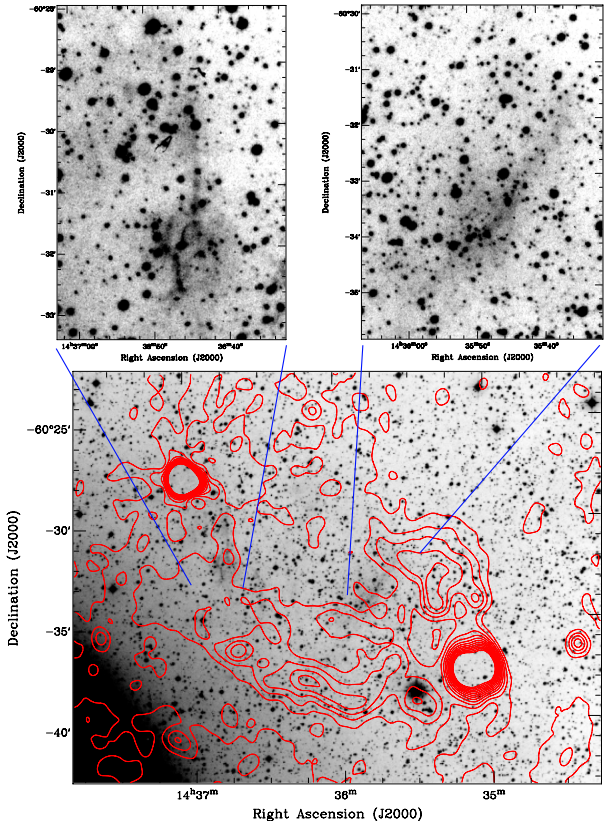

**Figure 14.** G315.4-0.3 in Hα overlaid with SUMSS 843 MHz radio contours from 0.004 to 0.03 Jy beam$^{-1}$. Several regions of extended Hα emission can be noticed. The upper two panels show enlarged areas for the two most prominent emission regions which show the Hα filaments more clearly. Both regions follow the inner boundaries of the radio shell at 843 MHZ. The tight 843 MHz flux concentration to the south-west is a supposed HII region GAL 315.31-00.27 (Caswell & Haynes 1987) though no Hα emission is seen. The similar flux concentration in the upper left corner is an extragalactic radio source. The remaining portions of the SNR shell are non-thermal.

   Comparison of the 843 MHz radio data and IRAS 60 µm data in Whiteoak & Green (1996) clearly showed that G320.6-1.6 is a non-thermal source. However, given the large size of this complex area it would be real challenge to use optical spectra alone, even with integral field units, to properly disentangle the various and separate possible HII thermal regions from any extensive non-thermal emission from the SNR. However, selected, targeted spectra such as of the filamentary optical emission clearly seen over the G320.6-1.6 radio border on south-east side of remnant (see Fig. 16) could reveal any SNR origin. The status of the association of such optical filaments with the SNR remain uncertain.

**G332.4+0.1 - Kes 32.** Roger et al. (1985) described G332.4+0.1 or Kes 32 (MSH 16-51) as a young supernova remnant (age ~5000 years) with a jet on the north side (see Fig. 17 which shows the radio contours "breaking-out" from a uniform shell structure to the north) and at a distance of some 5 kpc (on average). There is some faint diffuse emission in the region of the radio jet, but on the east side we found a more distinct chain of two broken, nebulous filaments ex- tending some 8 arcmin in a north-south direction (enlarged plate on Fig. 17). The SUMMS radio contours well-follow the most northerly component of these filaments but the most intense radio emission of 0.7 Jy beam$^{-1}$ is cetred between the two filament groups while there does not appear to be any radio emission emanating from the region of the southerly filament group. In Fig. 17 the lowest contours were set at 0.04 Jy beam$^{-1}$ to clearly show position of the under- lying optical emission although the radio surface brightness goes as low as 0.01 Jy beam$^{-1}$.

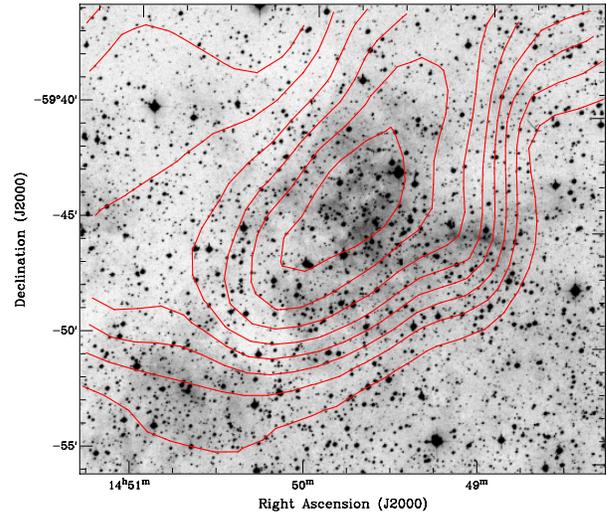

**Figure 15.** SHS Hα image of Galactic SNR G317.3-0.2 overlaid with 0.1 to 1.27 Jy beam$^{-1}$ PMN 4850 MHz radio contours. There is an excellent match between the radio and Hα emissions in the central and south-west area of this SNR that may represent evidence for an optical counterpart.

**G332.5-5.6.** This object represents our first optical detection of this now confirmed remnant and we include it here for completeness but refer the reader to our paper Stupar et al. (2007b) where we published detailed optical, radio and X-ray observations of this remarkable Galactic SNR which we first confirmed based on its optical detection in the SHS (see also Discussion section regarding this SNR)

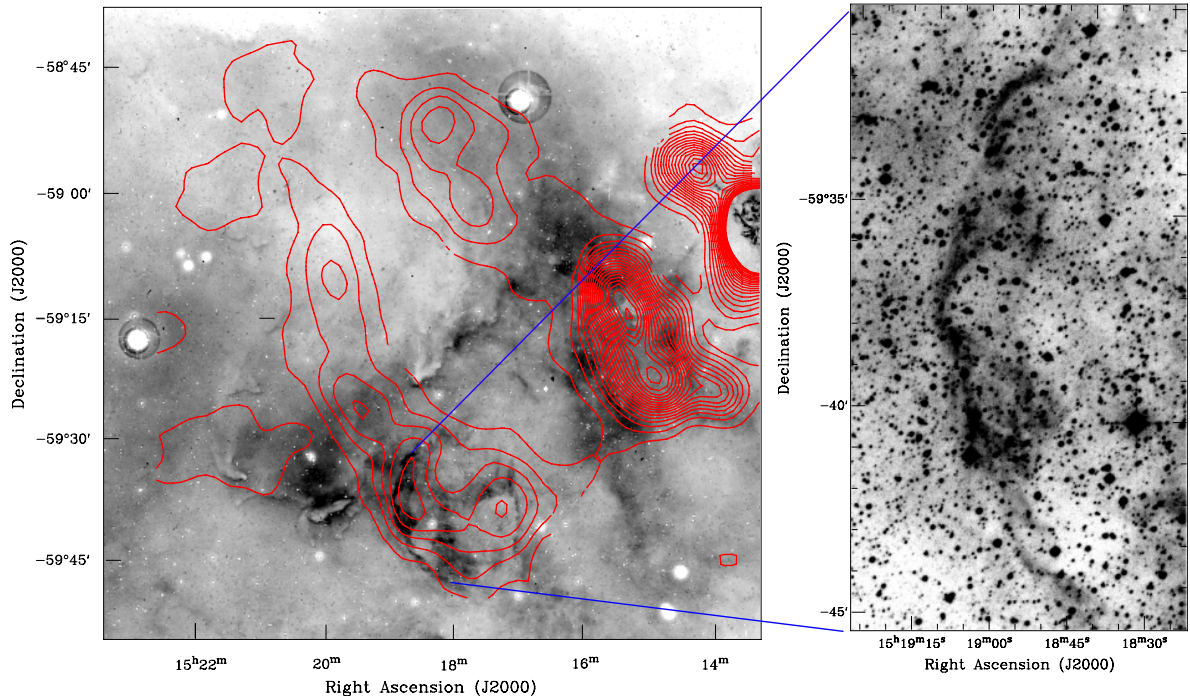

**Figure 16.** Hα quotient image (left) of G320.6-1.6 overlaid with PMN radio contours at 4850 MHz 0.02 to 1.7 Jy beam⁻¹. This is a very complicated region of optical emission and dusty regions. Nevertheless, inspection of our high resolution Hα images shows a probable mixture of emission from both a HII region and possibly also from the supernova remnant. The right hand side expanded image shows a high resolution Hα extract taken from the southern edge of the overall image of G320.6-1.6 which shows typical SNR filamentary optical structures matching for this area.

**G336.7+0.5.** There is extensive, general Galactic Hα emission to the north and all along the entire eastern side of this radio remnant which itself appears to fall in an area of increased extinction. Inside the PMN radio borders of this oval shaped region of radio emission associated with this SNR we found some 3 arcmin long enhanced Hα emission in an east-west direction to the south-west of the remnant core (see Fig. 18). At this stage it is not clear if this emission is associated with the SNR. We compared the crude radio morphology of G336.7+0.5 seen in the PMN with the higher resolution MOST catalogue image (see Whiteoak & Green 1996). There is excellent agreement in the overall oval shape between these two frequencies but the MOST data defines a clear radio shell with some internal structure what is shown on Fig. 19. The PMN shell appears shifted some 2 arcmin southwards where the match between the PMN and strongest 843 MHz flux is located. The south-west edge of this shell appears to follow the observed Hα emission in this region and could well be associated though confirmatory optical spectra are required to confirm this.

**G350.0-2.0.** Gaensler (1998) observed this remnant with combined observations of VLA and Parkes at 1.4 GHz. It has been morphologically classified as a bilateral (or "barrel") SNR where the axis is parallel to the Galactic plane. The SNR's morphological form at this frequency is similar to that seen in the PMN at 4850 MHz and shown in Fig. 20. This work (see Gaensler 1998, and references therein) also supports the proposal that the area around this remnant actually presents three separate remnants: the bright north-west arc (seen also in Fig. 20), another arc lower in brightness east of this bright arc (not clearly defined in Fig. 20 due to the low PMN resolution) and a third remnant of circular form, again low in surface brightness but seen in the central part of the image as contours in Fig. 20. Interestingly, there is an area of approximately circular Hα emission structures at the centre of the overall radio structure shown in the enlargement of just - the Hα data to the top right of Fig. 20. The lower enlargement presents more typically filamentary Hα emission which follows the strong north-west radio arc at its southern extremity and could also be associated. These optical structures, which also appear to be pervaded by dust and areas of increased extinction, continue at lower intensity further along this radio arc. Their status should be clarified by planned follow-up optical spectroscopy.

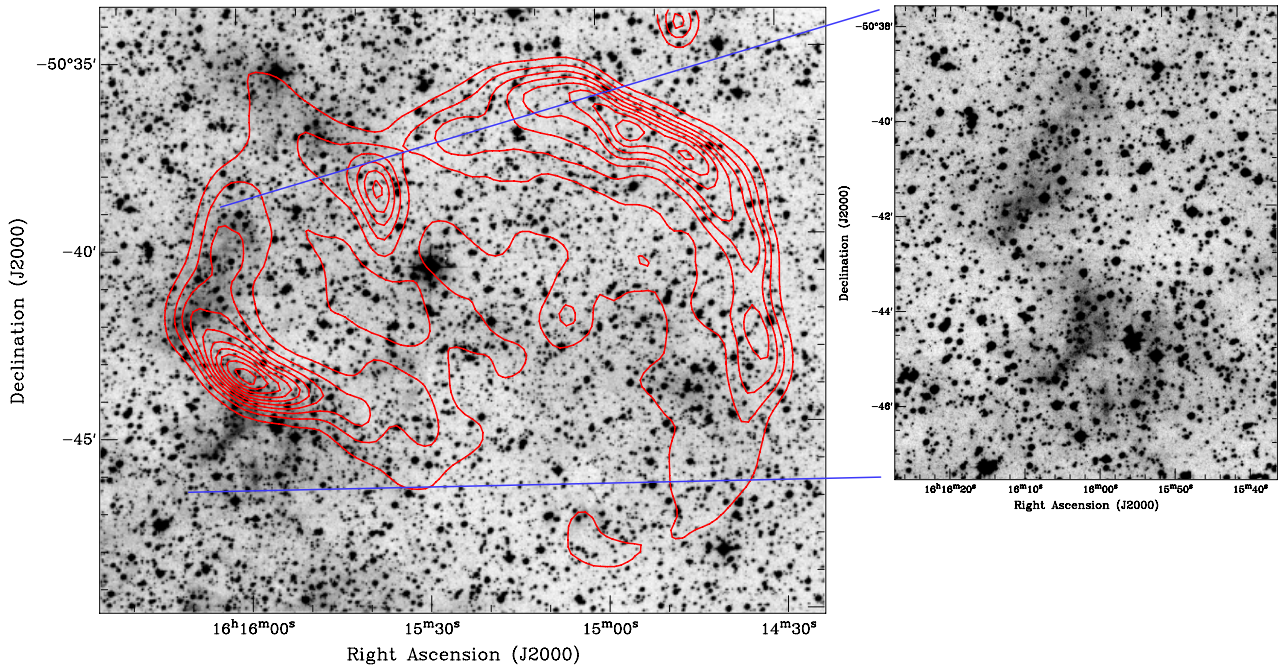

**Figure 17.** SUMSS 843 MHz contours from 0.04 to 0.7 Jy beam$^{-1}$ show the general shell-like form of SNR G332.4+0.1 (or Kes 32) but with enhanced opposing radio arcs to the south-east and north-west with contours from 0.01 Jy beam$^{-1}$ (see text). The right figure shows an enlarged region centred on the 10 arcmin long, prominent Hα filamentary emission to the east side of the remnant.

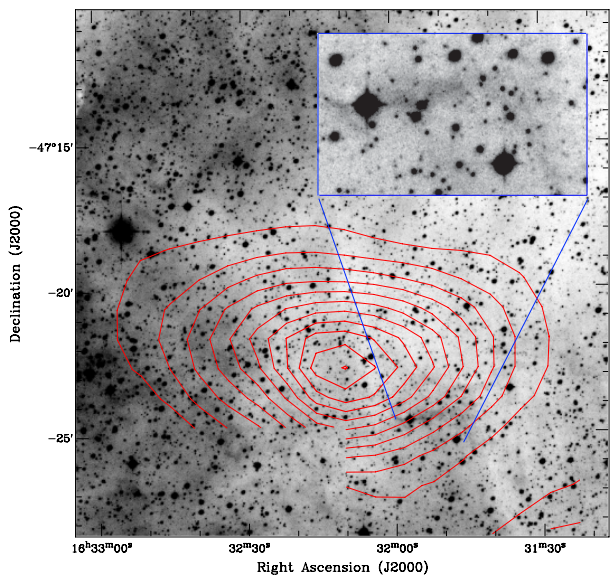

**Figure 18.** Hα SHS image of SNR G336.7+0.5 with an inset window showing a region of increased Hα emission extending ∼ 2 arcmin east-west. The PMN radio contours are from 0.02 to 0.52 Jy beam$^{-1}$ and reveal a compact, resolved, oval source with no internal structure discernable at the low PMN resolution.

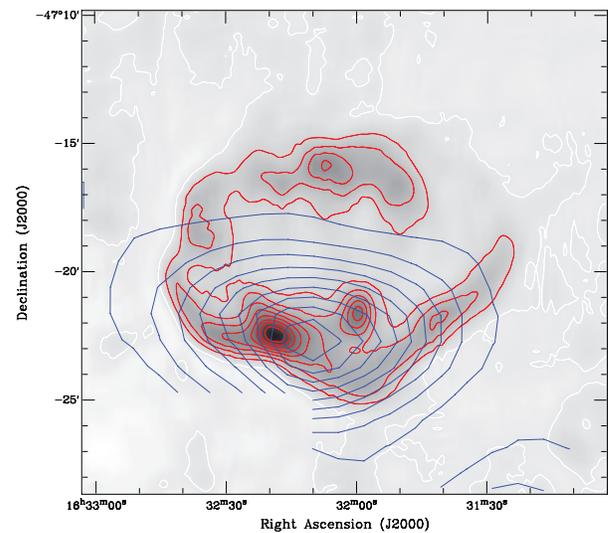

**Figure 19.** SNR G336.7+0.5 from the MOST Catalogue (Whiteoak & Green 1996) with contours (red) going from 0.028 to 0.25 Jy beam$^{-1}$ overlaid with the same image (blue contours) of this remnant from Fig. 18 at 4850 MHz. Shift in overall position between two frequencies is clear although they are matched in the area where the radio flux is strongest on both images. One can notice missing of the radio flux on the south-east side of PMN 4850 MHz image probably due to solar interference during the scanning process.

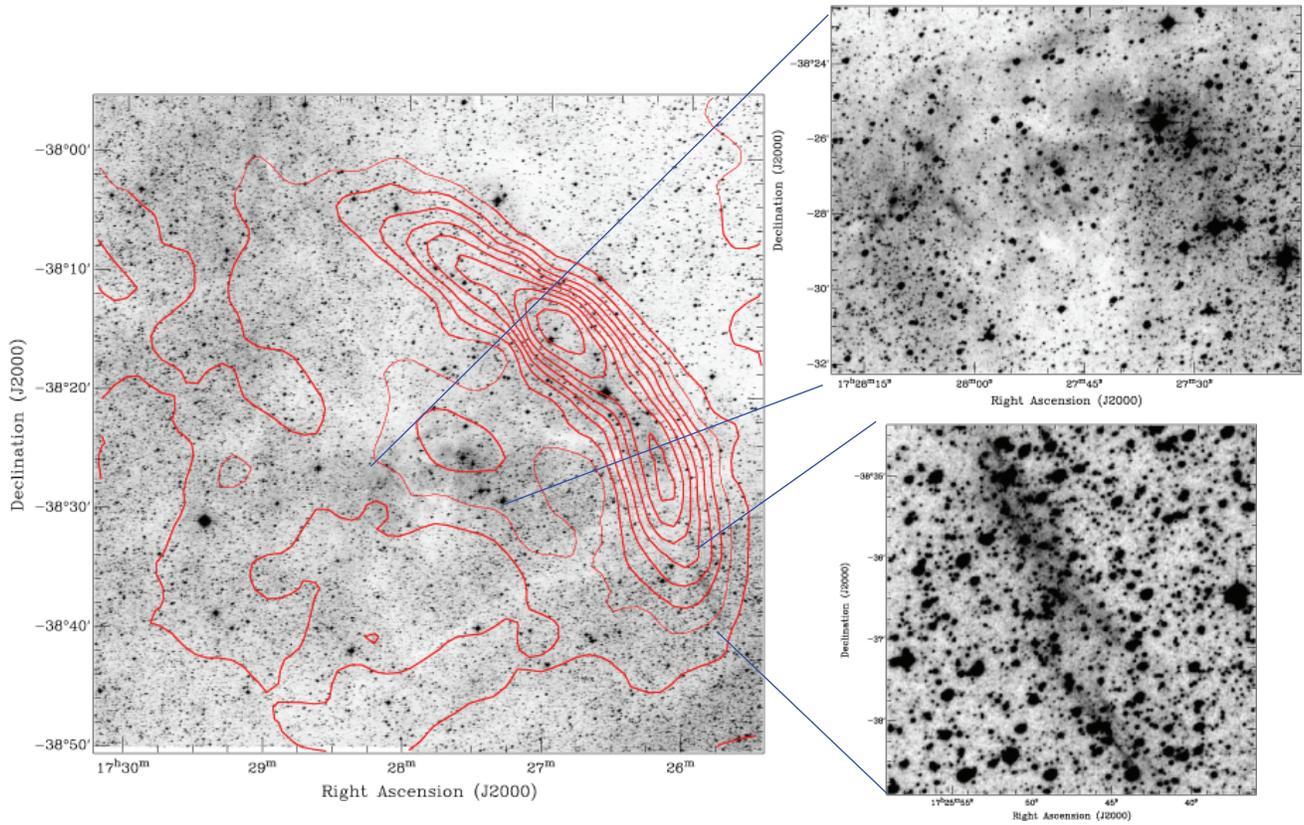

**Figure 20.** SNR G350.0-2.0 shown as an SHS Hα image with overlaid PMN contours going from 0.01 1 to 0.5 Jy beam⁻¹. The radio data is dominated by a powerful 40 arcmin emission arc to the north west. The two enlarged Hα images to the right focus on two areas of intense Hα emission in the form of fractured emission clouds found near the centre of the radio remnant (top) and some elongated filaments at the southern extent of the prominent radio arc (bottom) which we suggest could be associated with the SNR(s).

**G359.1-0.5.** This is a classic shell-like radio detected SNR towards the Galactic centre which appears to have an embedded approximately circular region of Hα emission within the shell. This is clearly seen in Fig. 22 which shows the radio shell and the internal Hα flocculent cloud slightly to the north-west side of the radio shell, offset by ~4 arcmin from the centre. It is some ~ 12 arcmin in size from the north-west to south-east and has the highest surface brightness to the south. Deeper analysis shows some kind of "outflow" from the cloud to the north-west. The cloud itself is rather inhomogeneous and consists of emission blobs and wider filaments. If this structure was found outside of the radio border of SNR, we would be tempted to classify it as new optical remnant in its own right similar to G243.9+9.8 in Stupar, Parker & Filipovic (2008) or to the well known SNR IC 443 (see images in Lozinskaia 1979; Fesen 1984).

Unfortunately, the PMN survey does not cover this area of the sky so we could not compare the Hα data with a radio image at 4850 MHz. The NVSS at 1.4 GHz shows just a few fragmented radio emission spots over the radio border seen in SUMSS at 843 MHz which covers most but not all of

G359.1-0.5 (see. Fig. 22) as this survey only covers up to δ=−30˚ and the northern part over this declination is not covered. As can be seen in Fig. 22 there appears to be no direct overlap between the 843 MHz radio flux map and the possible optical counterpart seen in Hα. This may simply be a chance alignment of a HII region and the SNR shell.

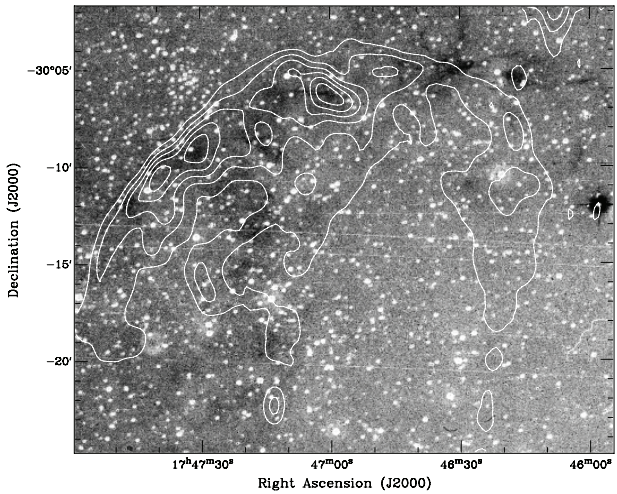

**Figure 21.** SUMSS 843 MHz contours of G359.0-0.9 situated in the Galactic centre reveal an incomplete radio shell. The radio contours from 0.03 to 0.26 Jy beam$^{-1}$ are overlaid over quotient image of Hα and short red showing on the north prominent Hα emission.

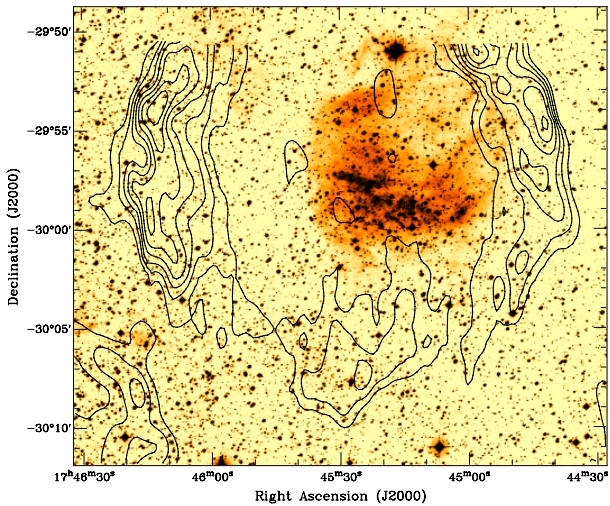

**Figure 22.** G359.1-0.5 is another SNR towards the Galactic centre. SUMSS 843 MHz contours from 0.01 to 0.22 Jy beam$^{-1}$ are overlaid on the Hα high resolution image. Note the striking, almost circular Hα cloud some 9.5 arcmin in diameter situated completely internal to the SNR shell but displaced to the north-west area of remnant by about 4 armin. Also note that the SUMSS radio image has no flux to the north due to the limited coverage (up to δ=-30°) of the survey. Optical spectroscopy across the optical emission is urgently needed to clarify this intriguing discovery.

## 4 DISCUSSION

Our Catalogue provides, for the first time, a compilation of imaging and positions of confirmed or suspected optical Hα emission newly associated with 24 Galactic SNRs. This represents a significant improvement of ~10% on top of the 20% of previously known Galactic SNRs with optical emission (out of 274; see Green 2009). Their uncovered Hα emissions predominantly appear in two distinct morphological forms: either as irregular or diffuse emission clouds or sets of isolated (sometimes chain-like) filamentary structures which are generally brighter than the emission clouds. Some of our new SNR Hα images demonstrate an excellent match between optical emission and regions of SNR radio flux seen at different frequencies (for specific examples see G11.1-1.0, G16.8-1.1 or G320.6-1.6). For some remnants the optical emission is irregular and fragmented over the radio detected area. This Hα emission is often matched with the strongest radio flux and mostly seen in typical filamentary form. These filaments represent the shock front seen face- on, although the shock is best seen when the front is on the boundary of the remnant seen edge-on (with enhanced limb brightening). This does not appear to be the case for any of 24 objects considered here.

Most of the catalogued objects exhibit strong Hα emission which is sufficient to undertake follow-up spectral analysis (see later discussion about bright filaments in the case of G332.5-5.6). After consideration of the apparent angular size and morphology of the sample we can conclude that, as expected, the observed remnants are mostly old. The optical emissions are, more or less, fragmented over their radio counterparts. Although there are cases where the Hα emission clearly partially follows the shape of the radio structure few objects show a full one-to-one correspondence across both the optical and radio morphological structures though SNR G15.1-1.6 and Kes 78 in particular exhibit excellent overall correspondence. Firm one-to-one optical-radio counterparts only really occur with young remnants. We do not have such cases among the observed sample.

We suspect that most of remnants in our sample have evolved toward the last, dissipation phase and that they are probably somewhere in the snowplow phase not having quite reached the last dissipation stage when they dissolve into the ISM. When these SNRs finally mix with the ISM this also means that the local magnetic field comes up to the level of that of the Galactic background. In such instances non- thermal emission is uncertain and radio detection is often not possible. This is certainly not the case for our sample where all objects have previously detected radio emission.

In instances where the radio observations show a mixture of thermal and no-thermal emission (some examples are included in this compilation) optical emission line spectra of the optical counterparts can provide useful diagnostic power in providing for an overall definition of some remnants due to the much better 1 arcsecond resolution of the optical observations compared to their radio counterparts. Similarly, the high definition morphological optical images of these remnants compared with mid-infrared emission at 8.3 μm from the MSX satellite mission (see Price et al. 2001) can help us to precisely define the thermal emission from contaminating HII regions and also reveal the difference in position of the thermal and non-thermal emission components. In another words, we can use this comparison to confirm if an object is a HII region (where we can detect image fluorescent emission from polycyclic aromatic hydrocarbons (or PAH) - see details in Price et al. (2001)) or non-thermal emission from an SNR where PAHs are not registered (see the similar relationship between PAH and radio emission in Cohen & Green (2001)). Sometimes, association of an object in the optical (and radio) with surrounding objects of known distances, compared with mid-infrared emission at 8.3 μm from MSX, can help in distance determination (e.g. the case of SNR G213.0-0.6 - Stupar & Parker in preparation).

Note that the optical spectra of shocks from SNRs are predominantly radiative (with some Balmer line emission but also strong [N II], [S II], [O I] and other forbidden lines) but it is possible that spectra of some objects from the sample will show non-radiative (or Balmer-dominated) emission. It is true that non-radiative emission is mostly connected with younger remnants (strong hydrogen lines, high speed shocks) but Gerardy & Fesen (2007) showed that this kind of spectra can also be obtained in some older remnants. Hence, there is possibility that some remnants from this Catalogue may exhibit non-radiative spectra. Standard emission line diagnostics such as used by Fesen, Blair & Kirshner (1985) and recently reported by Frew & Parker (2010) will be used to elucidate the status of these optical counterparts and to identify contaminants.

On the other hand, radiative spectra, particularly the ratio of [S II] 6717/6731 Å, can help in determining the age of remnants as velocity and electron density estimates differ between young and old remnants. Furthermore, recent optical spectral observations of SNRs have been used to show that synchrotron optical emission can be observed in- side SNR shocks. Actually, Jiang et al. (2010), predicted that the lower bound of synchrotron optical emission from SNR shocks should be $1.53 \times 10^{-11}$ erg s$^{-1}$ cm$^{-2}$ . They used our spectra from SNR G332.5-5.6 (object from this Catalogue; see Stupar et al. 2007b) and show that our flux of $10^{-12}$ erg s$^{-1}$ cm$^{-2}$ was actually very close to that predicted, giving us further confidence in our techniques.

Figure 23 shows the distribution of all Galactic remnants with a recognised optical counterpart, where data (Galactic coordinates) of previous optical observations are marked with a red symbol (taken from Green 2009) while our proposed new optical detections are in blue. Our sample only covers the Southern Galactic plane which restricts our latitude range but it is clear that we follow the general trend of previous optical detections of remnants close to the Galactic plane. This is expected given the concentration of these young objects to the mid-plane. However, we have found one SNR G332.5-5.6 with Hα emission with a latitude as high as $b = -5.6$. Also, due to heavy extinction, it has not been common to detect optical emission from known SNRs around the Galactic centre where only radio observations of Galactic SNRs are usually possible. However, in this region (less than 1° around Galactic centre) for the first time we detected Hα emission (see locations on Fig. 23) inside the radio borders of G359.0-0.9 and G359.1-0.5. Yet again, this shows the superiority of the SHS over previous broad and narrow-band optical sky surveys in terms of resolution and sensitivity. Future Hα surveys such as VPHAS+ will soon begin with the VST, which have better resolution and above all higher sensitivity, will be very well placed to register many more optical emission structures associated with known SNRs close to Galactic centre and towards the mid plane ($|b| \leq 5$ degrees).

# 5 ACKNOWLEDGEMENTS

We are thankful to the staff of the Wide Field Astronomy Unit at the Royal Observatory Edinburgh for their help dur- ing the visual inspection of original films of the AAO/UKST Hα Survey of the southern Galactic plane. We would also like to thank the reviewer for constructive comments that have improved the paper.

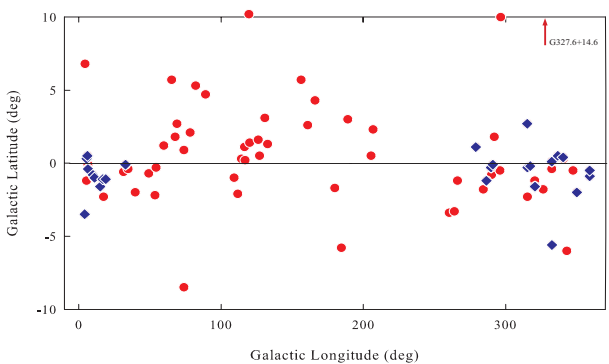

**Figure 23.** Distribution of optically detected known Galactic SNRs. Coordinates are taken from Green (2009). Red spots give the distribution of previously detected remnants, and blue dots are from this work. One can notice our Hα detection for the two remnants (G359.0-0.9 and G359.1-0.5) in the area close to Galactic centre, not expected due to high concentration of dust in this area.